\documentclass[12pt,thmsa]{article}
\usepackage{amssymb}

\usepackage{sw20aip}

\input{tcilatex}
\begin{document}

\author{E. Elizalde$^{1,2}$, E. J. Ferrer$^{1,3}$ \and and V. de la Incera$^{2,3}$ \\
$^{1}${\small Institute for Space Studies of Catalonia, CSIC, Edif. Nexus, }%
\\
{\small Gran Capita 2-4, 08034 Barcelona, Spain.} \\
$^{2}${\small University of Barcelona, Dept. of Structure and Constituents
of Matter,}\\
{\small Diagonal 647, 08028 Barcelona, Spain.}\\
$^{3}${\small Department of Physics, State University of New York at
Fredonia, }\\
{\small Houghton Hall 118, Fredonia, NY 14063, USA}}
\title{Neutrino Self-Energy and Index of Refraction in Strong Magnetic Field: A New
Approach}
\maketitle

\begin{abstract}
The Ritus' $E_{p}$ eigenfunction method is extended to the case of spin-1
charged particles in a constant electromagnetic field and used to calculate
the one-loop neutrino self-energy in the presence of a strong magnetic
field. From the obtained self-energy, the neutrino dispersion relation and
index of refraction in the magnetized vacuum are determined within the field
range $m_{e}^{2}\ll eB\ll M_{W}^{2}$. The propagation of neutrinos in the
magnetized vacuum is anisotropic due to the dependence of the index of
refraction on the angle between the directions of the neutrino momentum and
the external field. Possible cosmological implications of the results are
discussed.
\end{abstract}

\section{Introduction}

The main goal of this paper is to investigate the effects of magnetic fields
on neutrino propagation, a topic that has recently received increasing
attention. We are particularly interested in strong field effects. Its
possible application to astrophysics, where fields of the order of $10^{13}$ 
$G$, and even larger \cite{Astro}$,$ can be expected in supernova collapse
and neutron stars, makes this subject worth of detailed investigation. Just
to mention one of the several astrophysical applications of magnetic field
effects in neutrino physics, one may recall the suggestion that the
modification of the neutrino dispersion relation in a magnetized charged
medium \cite{Olivo}$^{,}$ \cite{Esposito}$^{,}$ \cite{Grasso} could serve to
explain the high velocity of pulsars \cite{Kusenko}$.$

Moreover, the presence of strong magnetic fields could have influenced the
propagation of neutrinos in the early Universe and have an imprint in
neutrino oscillations at those epochs\cite{Osc}$.$ The existence of
primordial magnetic fields (of the order of $10^{24}$ $G$ at the electroweak
scale) in the early Universe seems to be needed to explain the recent
observations of large-scale magnetic fields in a number of galaxies, in
galactic halos, and in clusters of galaxies \cite{Galaxies}$.$ These
primordial magnetic fields could be generated through different mechanisms,
as fluctuations during the inflationary universe \cite{Infla}$,$ at the GUT
scale \cite{GUT}$,$ or during the electroweak phase transition \cite{EW}$,$
among others.

Calculations of neutrino self-energies taking into account non-perturbative
effects of magnetic fields have been carried out in several works \cite
{Grasso}$^{,}$ \cite{McKeon}$^{,}$ \cite{Feldman}$^{,}$ \cite{Valle}$,$
using the Schwinger method \cite{Schwinger}$.$ In the present paper, the
Ritus' technique, which was originally developed for the electron
self-energy in QED in the presence of electromagnetic backgrounds \cite
{Ritus}$^{,}$\cite{Ritus-Book}, is extended to the case of spin-1 charged
particles. The Ritus' method is based on a Fourier-like transformation that
diagonalizes in the momentum space the Green's functions of the charged
particles in the presence of a constant magnetic field. This approach \cite
{Ritus}$^{,}$\cite{Ritus-Book} is particularly convenient for the strong
field case, where one can constraint the calculation to the contribution of
the lower Landau level (LLL).

In this work we calculate the vacuum (zero temperature ($T=0$), zero density
($\mu =0$)) contribution of the neutrino dispersion relation at strong
magnetic field ($m_{e}^{2}\ll eB\ll M_{W}^{2}$, $\ m_{e}$ is the electron
mass and $M_{W}$ is the W-boson mass). As discussed below, such a strong
field can be expected to exist in the neutrino decoupling era. One of our
main results is the existence of an anisotropic propagation of neutrinos in
the strong magnetic field, even in the absence of a medium ($\mu =0$). The
anisotropy is due to the dependence of the index of refraction on the angle
between the directions of the neutrino momentum and the external field. We
also find that the terms explicitly depending on the mass of the charged
lepton are negligible small (of order $1/M_{W}^{4}$), while the leading term
results of order $1/M_{W}^{2}$, thus rather significant.

The plan of the paper is as follows. In Section II, for the sake of
understanding and completeness, we review the Ritus' method for the Green's
function of spin-1/2 particles in the presence of a constant magnetic field.
Then, we extend this method to the spin-1 charged particle case in the
background of a constant magnetic field (corresponding to the crossed
electromagnetic field ($\mathbf{E\cdot B}=0$) case). In Section III, we use
the results of Section II to calculate, in momentum space, the one-loop
neutrino self-energy in the presence of a constant magnetic field. The
neutrino dispersion relation and index of refraction are obtained in Section
IV in the strong-field approximation ($m_{e}^{2}\ll eB\ll M_{W}^{2}$). In
Section V, we make our final remarks and discuss a possible cosmological
realization of the adopted strong-field approximation.

\section{Green's Functions at $B\neq 0$ in the Momentum Representation}

The diagonalization, structure and properties of the Green's functions of
the electron and photon in an intense magnetic field were considered exactly
in external and radiative fields by Ritus in Refs. [15] and [16]. Ritus'
formulation provides an alternative method to the Schwinger approach to
address QFT problems on electromagnetic backgrounds. In Ritus' approach the
transformation to momentum space of the spin-1/2 particle Green's function
in the presence of a constant magnetic field is carried out using the $%
E_{p}(x)$ functions\cite{Ritus}$^{,}$\cite{Ritus-Book}$,$ corresponding to
the eigenfunctions of the spin-1/2 charged particles in the electromagnetic
background. The $E_{p}(x)$ functions plays the role, in the presence of
magnetic fields, of the usual Fourier $e^{ipx}$ functions in the free case.
This method is very convenient for strong-field calculations, where the LLL
approximation is plausible and for finite temperature calculations.

In this section we will extend the Ritus' method to the case of spin-1
charged particles. This extension will allow us to obtain a diagonal in
momentum space Green's function for the spin-1 charged particle in the
presence of a constant magnetic field. Our results are important to
investigate the behavior of charged W-bosons in strong magnetic fields.

\subsection{Electron Green's function}

For the sake of understanding, we summarize below the results obtained by
Ritus\cite{Ritus}$^{,}$\cite{Ritus-Book} for the spin-1/2 case. The Green's
function equation for the spin-1/2 particle in the presence of a constant
electromagnetic field without radiative corrections is given by

\begin{equation}
\left( \gamma .\Pi +m_{e}\right) S(x,y)=\delta ^{(4)}(x-y)  \label{0}
\end{equation}
where

\begin{equation}
\Pi _{\mu }=-i\partial _{\mu }-eA_{\mu }^{ext},\qquad \mu =0,1,2,3  \label{2}
\end{equation}

Taking into account that

\begin{equation}
\left[ S(x,y),\left( \gamma .\Pi \right) ^{2}\right] =0  \label{1}
\end{equation}
it follows that $S(x,y)$ will be diagonal in the basis spanned by the
eigenfunctions of $\left( \gamma .\Pi \right) ^{2}$

\begin{equation}
\left( \gamma .\Pi \right) ^{2}\Psi _{p}\left( x\right) =-p^{2}\Psi
_{p}\left( x\right)  \label{3}
\end{equation}
Since $\left[ \left( \gamma .\Pi \right) ^{2},\Sigma _{3}\right] =\left[
\left( \gamma .\Pi \right) ^{2},\gamma _{5}\right] =\left[ \Sigma
_{3},\gamma _{5}\right] =0$, one can easily find the eigenfunctions $\Psi
_{p}$ in the chiral representation, where $\Sigma _{3}=i\gamma _{1}\gamma
_{2}$ and $\gamma _{5}$ are both diagonal and have eigenvalues $\sigma =\pm
1 $ and $\chi =\pm 1$, respectively. The eigenfunctions are given by

\begin{equation}
\Psi _{p}\left( x\right) =E_{p\sigma \chi }(x)\Theta _{\sigma \chi }
\label{4}
\end{equation}
with $\Theta _{\sigma \chi }$ the bispinors which are the eigenvectors of $%
\Sigma _{3}$ and $\gamma _{5}$. We are considering the metric $g_{\mu \nu
}=diag(-1,1,1,1)$.

In the crossed field case ($\mathbf{E}\cdot \mathbf{B}=0$), one can always
select the potential in the Landau gauge

\begin{equation}
A_{\mu }^{\mathit{ext}}=Bx_{1}\delta _{\mu 2}  \label{5}
\end{equation}
which corresponds to a constant magnetic field of strength $B$ directed
along the $z$ direction in the rest frame of the system. The fermion
eigenfunctions are then given by the combination

\begin{equation}
E_{p}(x)=\sum\limits_{\sigma }E_{p\sigma }(x)\Delta (\sigma ),  \label{6}
\end{equation}
where 
\begin{equation}
\Delta (\sigma )=diag(\delta _{\sigma 1},\delta _{\sigma -1},\delta _{\sigma
1},\delta _{\sigma -1}),\qquad \sigma =\pm 1,  \label{7}
\end{equation}
and the $E_{p\sigma }$ functions are

\begin{equation}
E_{p\sigma }(x)=N(n)e^{i(p_{0}x^{0}+p_{2}x^{2}+p_{3}x^{3})}D_{n}(\rho )
\label{8}
\end{equation}
In Eq. (\ref{8}), $N(n)=(4\pi \left| eB\right| )^{\frac{1}{4}}/\sqrt{n!}$ is
a normalization factor and $D_{n}(\rho )$ denotes the parabolic cylinder
functions\cite{handbook} with argument $\rho =\sqrt{2\left| eB\right| }%
(x_{1}-\frac{p_{2}}{eB})$ and positive integer index

\begin{equation}
n=n(k,\sigma )\equiv l+\frac{\sigma }{2}-\frac{1}{2}\;\quad n=0,1,2,...,
\label{9}
\end{equation}
The integer $l$ in Eq. (\ref{9}) labels the Landau levels. In a pure
magnetic background the $\chi $ dependence of the eigenfunctions $E_{p\sigma
\chi }$ drops away.

The $E_{p}$ functions satisfy

\begin{equation}
\gamma .\Pi E_{p}(x)=E_{p}(x)\gamma .\overline{p}  \label{10}
\end{equation}
where

\begin{equation}
\overline{p}_{\mu }=\left( p_{0},0,-sgn\left( eB\right) \sqrt{2\left|
eB\right| l},p_{3}\right)  \label{11}
\end{equation}

One can easily check that these functions are both orthonormal

\begin{equation}
\int d^{4}x\overline{E}_{p^{\prime }}(x)E_{p}(x)=(2\pi )^{4}\widehat{\delta }%
^{(4)}(p-p^{\prime })\equiv (2\pi )^{4}\delta _{kk^{\prime }}\delta
(p_{0}-p_{0}^{\prime })\delta (p_{2}-p_{2}^{\prime })\delta
(p_{3}-p_{3}^{\prime })  \label{12}
\end{equation}
and complete

\begin{equation}
\sum_k\hspace{-0.5cm}\int%
d^{4}pE_{p}(x)\overline{E}_{p}(y)=(2\pi )^{4}\delta ^{(4)}(x-y)  \label{13}
\end{equation}
Here we have used the notation $\overline{E}_{p}(x)\equiv \gamma
^{0}E_{p}^{\dagger }\gamma ^{0}$ and $%
\sum\hspace{-0.4cm}\int%
d^{4}p=\sum\limits_{k}\int dp_{0}dp_{2}dp_{3}.$

Using the functions $E_{p}$ as a new basis, we obtain, thank to the
properties (\ref{10}) and (\ref{13}), a representation of the fermion
Green's function in the presence of a constant magnetic field which is
diagonal in $p$

\[
S(p,p^{\prime })\equiv \int d^{4}xd^{4}y\overline{E}_{p}(x)S(x,y)E_{p^{%
\prime }}(y) 
\]

\begin{equation}
=(2\pi )^{4}\widehat{\delta }^{(4)}(p-p^{\prime })\frac{1}{\gamma .\overline{%
p}+m_{e}}  \label{14}
\end{equation}

The main idea of the Ritus' approach is, therefore, to use the
eigenfunctions $E_{p}(x)$, which correspond to the asymptotic states of the
particles in the presence of a constant external electromagnetic field, to
perform a Fourier-like transformation that diagonalizes the Green's
functions in the momentum space. The advantage of the representation (\ref
{14}) is that the Green's function is simply given in terms of the
eigenvalues (\ref{11}).

\subsection{W-Boson Green's function}

Let us consider now the electroweak theory in the presence of a constant
magnetic field corresponding to the potential (\ref{5}). We choose the
following gauge conditions

\begin{equation}
F_{A}=\partial ^{\mu }A_{\mu }  \label{14a}
\end{equation}

\begin{equation}
F_{z}=\partial ^{\mu }Z_{\mu }+\alpha _{z}M_{z}\phi _{3}  \label{14b}
\end{equation}

\begin{equation}
F_{W}^{+}=D^{\mu }W_{\mu }^{+}+i\alpha _{W}M_{W}\phi  \label{14c}
\end{equation}

\begin{equation}
F_{W}^{-}=D^{*\mu }W_{\mu }^{-}-i\alpha _{W}M_{W}\phi ^{*}  \label{14d}
\end{equation}
with

\begin{equation}
D_{\mu }=\partial _{\mu }-ieA_{\mu }^{\mathit{ext}}  \label{14e}
\end{equation}
In the above expressions the customary notation \cite{ChengBook} for the
electroweak fields is used.

With the gauge conditions (\ref{14a})-(\ref{14d}) the Green's function
equation for the W-boson takes the form

\begin{equation}
\left[ \left( \Pi ^{2}+M_{W}^{2}\right) \delta _{\nu }^{\mu }-2ieF^{\mu
}\,_{\nu }+(\frac{1}{\alpha _{W}}-1)\Pi ^{\mu }\Pi _{\nu }\right] G_{\mu
}\,^{\nu }(x,y)=\delta ^{\left( 4\right) }(x,y)  \label{15}
\end{equation}
where $\Pi _{\mu }$ is given in Eq. (\ref{2}).

To solve Eq. (\ref{15}) it is convenient to perform first a rotation in the
Lorentz space using the transformation matrix

\begin{equation}
P^{\mu }\,_{\alpha }=\frac{1}{\sqrt{2}}\left( 
\begin{array}{llll}
\sqrt{2} & 0 & 0 & 0 \\ 
0 & 1 & 1 & 0 \\ 
0 & i & -i & 0 \\ 
0 & 0 & 0 & \sqrt{2}
\end{array}
\right)  \label{16}
\end{equation}
which satisfies $P^{-1}=\widetilde{P}^{*}$, and the following relations

\begin{equation}
\Pi _{\mu }P^{\mu }\,_{\alpha }=\Pi _{\alpha }  \label{17}
\end{equation}

\begin{equation}
\left( P^{\alpha }\;_{\mu }\right) ^{-1}\left[ iF^{\mu }\;_{\nu }\right]
P^{\nu }\,_{\beta }=-B\left( S_{3}\right) ^{\alpha }\,_{\beta }  \label{18}
\end{equation}

In the above expressions the notation

\begin{equation}
\Pi _{\alpha }=\left( \Pi _{0},\Pi _{+},\Pi _{-},\Pi _{3}\right)  \label{19}
\end{equation}

\begin{equation}
\Pi _{\pm }=\left( \Pi _{1}\pm i\Pi _{2}\right) /\sqrt{2}  \label{20}
\end{equation}
was introduced. $S_{3}$ represents the diagonal spin-one matrix

\begin{equation}
S_{3}=diag(0,1,-1,0).  \label{21}
\end{equation}

After doing the rotation (\ref{16}), the Green's function equation (\ref{15}%
) can be written as

\begin{equation}
\left[ \left( \Pi ^{2}+M_{W}^{2}\right) \delta _{\beta }^{\alpha }+2eB\left(
S_{3}\right) ^{\alpha }\,_{\beta }+(\frac{1}{\alpha _{W}}-1)\Pi ^{\alpha
}\Pi _{\beta }\right] G_{\alpha }\,^{\beta }(x,y)=\delta ^{\left( 4\right)
}(x,y)  \label{22}
\end{equation}

Now we can use the Feynman gauge\cite{footnote1}$,$ $\alpha _{w}=1$ , and
follow an approach similar to the spin-1/2 case in order to find a diagonal
in $p$ solution of Eq. (\ref{22}).

We start by solving the eigenvalue equation

\begin{equation}
\widehat{D}^{\alpha }\,_{\beta }\Phi _{k}^{\beta }\left( x\right) =\overline{%
k}^{2}\Phi _{k}^{\alpha }\left( x\right)  \label{23}
\end{equation}
where

\begin{equation}
\widehat{D}^{\alpha }\,_{\beta }=\left( \Pi ^{2}+2eBS_{3}\right) ^{\alpha
}\,_{\beta }  \label{24}
\end{equation}
Because $\left[ \widehat{D},S_{3}\right] =0$, $\Phi _{k}^{\alpha }\left(
x\right) $ can be taken as a common eigenfunction to $\widehat{D}$ and $%
S_{3} $. The eigenvalue equation for $S_{3}$ is then given by

\begin{equation}
\left( S_{3}\right) ^{\alpha }\,_{\beta }\Phi _{k}^{\beta }\left( x\right)
=\eta \Phi _{k}^{\alpha }\left( x\right) ,\qquad \eta =0,\pm 1  \label{25}
\end{equation}
where $\eta $ denotes the different spin projections. From (\ref{25}) we can
write

\begin{equation}
\Phi _{k}^{\alpha }\left( x\right) =F_{k\eta }(x)\Bbb{E}_{\eta }^{\alpha },
\label{26}
\end{equation}
In Eq.(\ref{26}) $\Bbb{E}_{\eta }^{\alpha }$ represents the eigenfunctions
of $S_{3}$ corresponding to the eigenvalues $\eta =0,\pm 1$, in the
following way

\begin{equation}
\left( 
\begin{array}{l}
1 \\ 
0 \\ 
0 \\ 
0
\end{array}
\right) \text{ and }\left( 
\begin{array}{l}
0 \\ 
0 \\ 
0 \\ 
1
\end{array}
\right) \text{ for }\eta =0,\qquad \left( 
\begin{array}{l}
0 \\ 
1 \\ 
0 \\ 
0
\end{array}
\right) \text{ for }\eta =1,\;\text{and \ }\left( 
\begin{array}{l}
0 \\ 
0 \\ 
1 \\ 
0
\end{array}
\right) \text{ for }\eta =-1  \label{27}
\end{equation}
Note that there is a degeneracy for $\eta =0$.

The eigenvalue problem (\ref{23}) reduces now to find $F_{k\eta }(x)$ from
the differential equation

\begin{equation}
\left( \Pi ^{2}+2eB\eta -\overline{k}^{2}\right) F_{k\eta }(x)=0,\text{
\qquad }\eta =0,\pm 1  \label{28}
\end{equation}

From the definitions (\ref{19})-(\ref{20}) for the $\Pi $ operator in the
rotated system, and taking into account Eqs. (\ref{2}) and (\ref{5}), we can
propose

\begin{equation}
F_{k\eta }(x)=\exp \left( -ik_{0}x_{0}+ik_{2}x_{2}+ik_{3}x_{3}\right)
f_{k\eta }\left( x_{1}\right)  \label{29}
\end{equation}

Then, $f_{k\eta }\left( x_{1}\right) $ should satisfy

\begin{equation}
\left( \partial _{\xi }^{2}-\frac{\xi ^{2}}{4}+\varepsilon \right) f_{k\eta
}\left( \xi \right) =0  \label{30}
\end{equation}
where

\begin{equation}
\xi =\sqrt{2\left| eB\right| }\left( x_{1}+k_{2}/eB\right)  \label{31}
\end{equation}
and

\begin{equation}
\varepsilon =\frac{1}{2\left| eB\right| }\left( \overline{k}%
^{2}+k_{0}^{2}-k_{3}^{2}-2eB\eta \right)  \label{32}
\end{equation}

Eq. (\ref{30}) is the harmonic oscillator equation. Its physical solution
requires

\begin{equation}
f_{k\eta }(\xi )\rightarrow 0\text{ \ for \ }\xi \rightarrow \infty
\label{33}
\end{equation}

\begin{equation}
\varepsilon =n+1/2,\qquad n=0,1,2,...  \label{34}
\end{equation}

From the condition (\ref{34}) and the definition (\ref{32}), one has

\begin{equation}
\overline{k}^{2}=-k_{0}^{2}+k_{3}^{2}+2(n+1/2)eB+2\eta eB  \label{35}
\end{equation}

Considering the mass shell condition $\overline{k}^{2}=-M_{W}^{2}$ in Eq. (%
\ref{35}), we can write

\begin{equation}
k_{0}^{2}=\left( 2n+1\right) eB-g_{s}eB.S+k_{3}^{2}+M_{W}^{2}  \label{36}
\end{equation}
Eq. (\ref{36}) is the well-known energy-momentum relation for higher-spin
charged particles in interaction with a constant magnetic field\cite
{higher-spin}$.$ Here $g_{s}$ is the gyromagnetic radio of the particle with
spin $S$. For W bosons: $g_{s}=2$. In (\ref{36}) we can also identify the so
called\cite{Savvidy}$^{,}$\cite{Olesen} ``zero-mode problem'' at $%
eB>M_{W}^{2}$. As known, at those magnetic fields a vacuum instability
appears giving rise to a W-condensation\cite{Olesen}$.$ In our calculations
we restrict the magnitude of the magnetic field to $eB<M_{W}^{2}$, thus, no
tachyonic modes will be present.

From condition (\ref{35}), and taking into account that $\eta =0,\pm 1$, it
follows that

\begin{equation}
n=\mathit{m}-\eta -1,\qquad \mathit{m}=0,1,2,...  \label{36a}
\end{equation}
Then, from (\ref{35}) and (\ref{36a}) we can write

\begin{equation}
\overline{k}^{2}=-k_{0}^{2}+k_{3}^{2}+2(\mathit{m}-1/2)eB,\qquad \mathit{m}%
=0,1,2,...  \label{37}
\end{equation}
where $\mathit{m}$ are the Landau numbers of the energy spectrum of the W
bosons in the presence of the magnetic field.

The solution of Eq. (\ref{30}) satisfying the conditions (\ref{33}) and (\ref
{34}) is

\begin{equation}
f_{k\eta }(\xi )=2^{-n/2}\exp (-\xi ^{2}/4)H_{n}(\xi /2)  \label{38}
\end{equation}
where $H_{n}$ are Hermite polynomials. Finally, substituting this solution
into Eq. (\ref{29}) we can write

\begin{equation}
F_{k\eta }(x)=N(n)e^{i(p_{0}x^{0}+p_{2}x^{2}+p_{3}x^{3})}D_{n}(\xi )
\label{39}
\end{equation}
with $N(n)$ a normalization factor and $D_{n}(\xi )$ the parabolic cylinder
functions.

The functions (\ref{26}), together with the mass shell condition (\ref{36}),
respectively define the wave function and energy-momentum relation for
spin-1 particles in the presence of a constant crossed electromagnetic field
($\mathbf{E\cdot B}=0$). The study of the parallel-field case ($\mathbf{E}%
\mid \mid \mathbf{B}$) can be found in Ref. [23]$.$

Similarly to the spin-1/2 case (Eq. (\ref{6})), we can now form the
transformation matrix to momentum space for the W-boson Green's function,

\begin{equation}
\left[ F_{k}\left( x\right) \right] _{\alpha }\,^{\beta }=\sum\limits_{\eta
=0,\pm 1}F_{k\eta }(x)\left[ \Omega ^{(\eta )}\right] _{\alpha }\,^{\beta }
\label{40}
\end{equation}
where the basis matrices of the Lorentz space $\Omega ^{(\eta )}$ are
explicitly given by

\begin{equation}
\Omega ^{(\eta )}=diag(\delta _{\eta ,0},\delta _{\eta ,1},\delta _{\eta
,-1},\delta _{\eta ,0}),\qquad \eta =0,\pm 1  \label{41}
\end{equation}

Notice that the only difference between the $E_{p}(x)$ functions (Eq. (\ref
{6})) and the $F_{k}\left( x\right) $ functions (Eq. (\ref{40})) is given
through the basis of their matrix spaces. That is, for spin-1/2 the
transformation matrices are expanded in the spinorial basis $\Delta (\sigma
) $, while in the spin-1 case (Eq. (\ref{40})) they are expanded in the
Lorentz basis $\Omega ^{\eta }$.

It can be easily shown that the $F_{k}\left( x\right) $ are orthogonal

\begin{equation}
\int d^{4}xF_{k}\left( x\right) F_{k^{\prime }}^{*}\left( x\right) =(2\pi
)^{4}\widehat{\delta }^{(4)}(k-k^{\prime })  \label{42}
\end{equation}
with normalization factor given by $N^{2}=\sqrt{4\pi \left| eB\right| }/n!$;
and complete

\begin{equation}
\sum_m\hspace{-0.5cm}\int%
d^{4}kF_{k}\left( x\right) F_{k}^{*}(y)=(2\pi )^{4}\delta ^{(4)}(x-y)
\label{43}
\end{equation}

Using the completeness property (\ref{43}), one can prove that the Green's
function

\begin{equation}
G_{F}(x,y)_{\alpha }\,^{\beta }=%
\sum_m\hspace{-0.5cm}\int%
\frac{d^{4}k}{\left( 2\pi \right) ^{4}}F_{k}\left( x\right) \frac{\delta
_{\alpha }\,^{\beta }}{\overline{k}^{2}+M_{W}^{2}}F_{k}^{*}(y)  \label{44}
\end{equation}
is a solution of Eq. (\ref{22}) in the Feynman gauge.

We can use the matrix $P$ (\ref{16}) to perform a similarity transformation
of the Green's function (\ref{44}) in order to represent it in the
rectangular Lorentz space as

\begin{equation}
G_{F}(x,y)_{\mu }\,^{\nu }=%
\sum_m\hspace{-0.5cm}\int%
\frac{d^{4}k}{\left( 2\pi \right) ^{4}}\Gamma _{k}^{\alpha }\,_{\mu }\left(
x\right) \frac{\delta _{\alpha }\,^{\beta }}{\overline{k}^{2}+M_{W}^{2}}%
\Gamma _{\,k\beta }^{\dagger }\,^{\nu }(y)  \label{45}
\end{equation}
where

\[
\Gamma _{k}^{\alpha }\,_{\mu }\left( x\right) =P^{\alpha }\,_{\gamma
}F_{k}\left( x\right) P^{-1\gamma }\,_{\mu } 
\]

\begin{equation}
=\frac{1}{2}\left( 
\begin{array}{llll}
2\mathcal{H}_{0}\left( x\right) & 0 & 0 & 0 \\ 
0 & \mathcal{H}_{+}\left( x\right) & -i\mathcal{H}_{-}\left( x\right) & 0 \\ 
0 & i\mathcal{H}_{-}\left( x\right) & \mathcal{H}_{+}\left( x\right) & 0 \\ 
0 & 0 & 0 & 2\mathcal{H}_{0}\left( x\right)
\end{array}
\right)  \label{46}
\end{equation}
and the $\mathcal{H}_{0}$, $\mathcal{H}_{\pm }$ functions are given by

\begin{equation}
\mathcal{H}_{0}\left( x\right) =\mathcal{F}_{\mathit{m}-1}\left( x\right)
,\quad \mathcal{H}_{\pm }\left( x\right) =\mathcal{F}_{\mathit{m}-2}\left(
x\right) \pm \mathcal{F}_{\mathit{m}}\left( x\right)  \label{47}
\end{equation}
In Eq. (\ref{47}) $\mathcal{F}_{n}\left( x\right) $ represents the functions 
$F_{k\eta }(x)$ evaluated at the different spin projections $\eta =0,\pm 1$.
That is, using the relation (\ref{36a}) and evaluating $\eta $ on each spin
projection, we obtain the different values of $n$ appearing as subindexes of 
$\mathcal{F}_{n}\left( x\right) $ in terms of the Landau levels $\mathit{m}$.

From Eq. (\ref{45}) and the orthogonality condition (\ref{42}), the W-boson
Green's function in a constant magnetic field can be written in the Lorentz
rectangular frame as the following diagonal function of momenta

\begin{equation}
G_{F}(k,k^{\prime })_{\mu }\,^{\nu }=\left( 2\pi \right) ^{4}\widehat{\delta 
}^{(4)}(k-k^{\prime })\frac{\delta _{\mu }\,^{\nu }}{\overline{k}%
^{2}+M_{W}^{2}}  \label{48}
\end{equation}
The eigenvalue $\overline{k}^{2}$ is given in Eq. (\ref{37}). We conclude
this section by stressing that the $\Gamma _{k}\left( x\right) $ matrices
play for the spin-1 particle Green's function the same role as the $E_{p}(x)$
matrices did for the spin-1/2 ones.

\section{Neutrino self-energy in the $E_{p}-\Gamma _{k}$ representation}

It is known\cite{Olivo}$^{,}$\cite{Grasso}$^{,}$\cite{Feldman}$^{,}$\cite
{Valle} that, to lowest order, the neutrino self-energy in a magnetic field
is given by the bubble diagram arising from the $e-W$ loop (See fig. 1).

\FRAME{ftbpFU}{2.4855in}{0.819in}{0pt}{\Qcb{The order-g$^{2}$ neutrino
bubble graph. Solid line corresponds to the charged lepton Green function in
a constant magnetic field and wiggly line corresponds to the W-boson Green
function in a constant magnetic field.}}{}{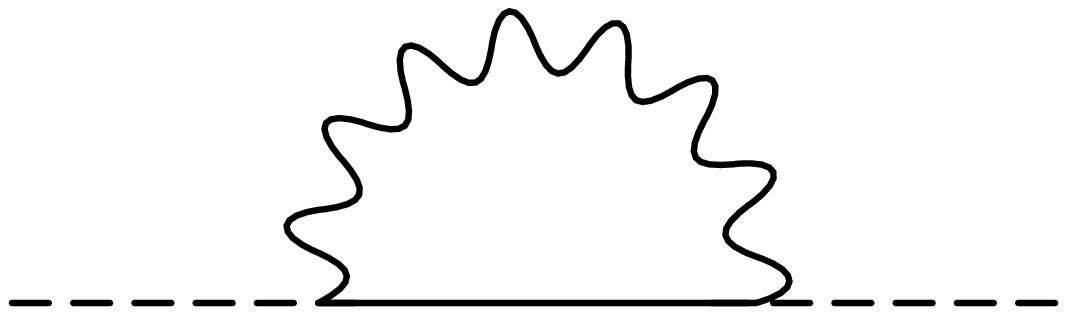}{\special{language
"Scientific Word";type "GRAPHIC";maintain-aspect-ratio TRUE;display
"PICT";valid_file "F";width 2.4855in;height 0.819in;depth 0pt;original-width
304.9375pt;original-height 99.1875pt;cropleft "0";croptop "1";cropright
"1";cropbottom "0";filename 'C:/selfen.ps';file-properties "XNPEU";}}This
diagram has been calculated using a perturbative expansion in the magnetic
field at $T\neq 0$ and $\mu \neq 0$ ($\mu $ is the chemical potential of
electrons) in Ref. [2]. Using the Schwinger proper-time method \cite
{Schwinger}$,$ which involves the magnetic interaction non-perturbatively,
the $e-W$ bubble has been calculated at $T\neq 0$ and $\mu \neq 0$ in Refs.
[13], and in vacuum (i.e. $T=0$ and $\mu =0$) in Refs. [11,12]. When
calculating the thermal contribution to the $e-W$ bubble with $\mu \neq 0$
or $\mu =0$ \cite{Olivo}$^{,}$\cite{Valle}$,$ the interaction between the
magnetic field and the charged W-bosons in the Green's function spectrum has
been often neglected under the assumption that $eB\ll M_{W}^{2}$ (in this
approximation the magnetic field does not appear in the poles of the W-boson
Green's function). The W-boson Green's function in the above mentioned
approximation, known in the literature as the ``contact approximation,''
takes the form

\begin{equation}
G_{0}^{\mu \nu }(x,y)\simeq \Phi (x,y)\int \frac{d^{4}k}{\left( 2\pi \right)
^{4}}e^{ik.(x-y)}\frac{g^{\mu \nu }}{M_{W}^{2}}  \label{49}
\end{equation}
where

\begin{equation}
\Phi (x,y)=\exp \left( i\frac{e}{2}y_{\mu }F^{\mu \nu }x_{\nu }\right)
\label{50}
\end{equation}
is the well known phase factor depending on the applied field\cite{Dittrich}$%
.$ That is, in the contact approximation the interaction of the magnetic
field with the W-bosons is restricted, in the W-boson Green's function, to
the phase factor (\ref{50}). The contact approximation must be carefully
handled in vacuum ($T=0$ and $\mu =0$), since it causes severe ultraviolet
divergences. To avoid those subtleties one can instead consider the
modification of the Green's function of the W-boson due to the external
magnetic field, on an equal foot with the electron, and then, only at the
end of the calculation take into account that the W-boson mass is the
largest scale in the problem. As shown below, this more careful approach
will prove to be convenient and useful for the calculation of the vacuum
contribution to the neutrino self-energy in the presence of a magnetic field.

\subsection{General formulation}

To calculate the neutrino self-energy in the one-loop approximation we start
from

\begin{equation}
\Sigma (x,y)=\frac{ig^{2}}{2}R\gamma _{\mu }S(x,y)\gamma ^{\nu
}G_{F}(x,y)_{\nu }\,^{\mu }L  \label{51}
\end{equation}
where $L,R=\frac{1}{2}(1\pm \gamma _{5})$, $G_{F}(x,y)_{\nu }\,^{\mu }$ is
the W-boson Green's function in the Feynman gauge (\ref{45}), and $S(x,y)$
is the electron Green's function (\ref{14}), that can be expressed in the
configuration space as

\begin{equation}
S(x,y)=%
\sum_{\it l}\hspace{-0.57cm}\int%
\frac{d^{4}q}{\left( 2\pi \right) ^{4}}E_{q}\left( x\right) \frac{1}{\gamma .%
\overline{q}+m_{e}}\overline{E}_{q}(y)  \label{52}
\end{equation}

Since the neutrino is an electrically neutral particle, the transformation
to momentum space of its self-energy can be carried out by the usual Fourier
transform

\begin{equation}
\left( 2\pi \right) ^{4}\delta ^{(4)}(p-p^{\prime })\Sigma (p)=\int
d^{4}xd^{4}ye^{-i(p.x-p^{\prime }.y)}\Sigma (x,y)  \label{53}
\end{equation}

Substituting with (\ref{51}), (\ref{45}) and (\ref{52}) in (\ref{53}) we
obtain

\[
\left( 2\pi \right) ^{4}\delta ^{(4)}(p-p^{\prime })\Sigma (p)=\frac{ig^{2}}{%
2}\int d^{4}xd^{4}ye^{-i(p.x-p^{\prime }.y)}\left\{ R\left[ \gamma _{\mu
}\left( 
\sum_{\it l}\hspace{-0.57cm}\int%
\frac{d^{4}q}{\left( 2\pi \right) ^{4}}E_{q}\left( x\right) \frac{1}{\gamma .%
\overline{q}+m_{e}}\overline{E}_{q}(y)\right) \right. \right. 
\]

\begin{equation}
\left. \left. \gamma ^{\nu }\left( 
\sum_m\hspace{-0.5cm}\int%
\frac{d^{4}k}{\left( 2\pi \right) ^{4}}\frac{\Gamma _{k}^{\alpha }\,_{\mu
}\left( x\right) \Gamma _{\,k\alpha }^{\dagger }\,^{\nu }(y)}{\overline{k}%
^{2}+M_{W}^{2}}\right) \right] L\right\}  \label{54}
\end{equation}

Taking into account that the spinor matrices (\ref{7}) satisfy the following
properties

\[
\Delta \left( \pm \right) ^{\dagger }=\Delta \left( \pm \right) ,\qquad
\Delta \left( \pm \right) \Delta \left( \pm \right) =\Delta \left( \pm
\right) ,\qquad \Delta \left( \pm \right) \Delta \left( \mp \right) =0 
\]

\[
\gamma ^{\shortparallel }\Delta \left( \pm \right) =\Delta \left( \pm
\right) \gamma ^{\shortparallel },\quad \gamma ^{\bot }\Delta \left( \pm
\right) =\Delta \left( \mp \right) \gamma ^{\bot }, 
\]

\begin{equation}
\quad L\Delta \left( \pm \right) =\Delta \left( \pm \right) L,\quad R\Delta
\left( \pm \right) =\Delta \left( \pm \right) R  \label{55}
\end{equation}
where the notation $\gamma ^{\shortparallel }=(\gamma ^{0},\gamma ^{3})$ and 
$\gamma ^{\bot }=(\gamma ^{1},\gamma ^{2})$ was introduced, and using the
definitions (\ref{46}), (\ref{47}), we obtain from (\ref{54})

\[
\left( 2\pi \right) ^{4}\delta ^{(4)}(p-p^{\prime })\Sigma (p)=\frac{-ig^{2}%
}{2}\int d^{4}xd^{4}y%
\sum_{\it l}\hspace{-0.57cm}\int%
\frac{d^{4}q}{\left( 2\pi \right) ^{4}}%
\sum_m\hspace{-0.5cm}\int%
\frac{d^{4}k}{\left( 2\pi \right) ^{4}}\frac{e^{-i(p.x-p^{\prime }.y)}}{(%
\overline{q}^{2}+m_{e}^{2})(\overline{k}^{2}+M_{W}^{2})} 
\]

\[
\left\{ 2\overline{q}_{\bot }\cdot \gamma ^{\bot }\left[ I_{\mathit{m}-1,%
\mathit{l}}(x)I_{\mathit{m}-1,\mathit{l}-1}^{*}(y)\Delta \left( -\right) +I_{%
\mathit{m}-1,\mathit{l}-1}(x)I_{\mathit{m}-1,\mathit{l}}^{*}(y)\Delta \left(
+\right) \right] \right. 
\]

\[
+\overline{q}_{\shortparallel }\cdot \gamma ^{\shortparallel }\left[ \left(
I_{\mathit{m}-2,\mathit{l}}(x)I_{\mathit{m}-2,\mathit{l}}^{*}(y)+I_{\mathit{m%
},\mathit{l}}(x)I_{\mathit{m},\mathit{l}}^{*}(y)\right) \Delta \left(
-\right) \right. 
\]

\[
\left. +\left( I_{\mathit{m}-2,\mathit{l}-1}(x)I_{\mathit{m}-2,\mathit{l}%
-1}^{*}(y)+I_{\mathit{m},\mathit{l}-1}(x)I_{\mathit{m},\mathit{l}%
-1}^{*}(y)\right) \Delta \left( +\right) \right] 
\]

\[
+\overline{q}_{\mu }\epsilon ^{1\mu 2\nu }\gamma _{\nu }\gamma ^{5}\left[
\left( I_{\mathit{m}-2,\mathit{l}}(x)I_{\mathit{m}-2,\mathit{l}}^{*}(y)-I_{%
\mathit{m},\mathit{l}}(x)I_{\mathit{m,l}}^{*}(y)\right) \Delta \left(
-\right) \right. 
\]

\begin{equation}
\left. \left. +\left( I_{\mathit{m}-2,\mathit{l}-1}(x)I_{\mathit{m}-2,%
\mathit{l}-1}^{*}(y)-I_{\mathit{m},\mathit{l}-1}(x)I_{\mathit{m},\mathit{l}%
-1}^{*}(y)\right) \Delta \left( +\right) \right] \right\} L  \label{56}
\end{equation}
In Eq. (\ref{56}) we used the compact notation

\begin{equation}
I_{\mathit{a},\mathit{b}}(x)=\mathcal{F}_{\mathit{a}}\left( x\right) E_{%
\mathit{b}}(x)  \label{57}
\end{equation}
with $\mathcal{F}_{\mathit{a}}\left( x\right) =F_{k\eta }(x)$ and $E_{%
\mathit{b}}(x)=E_{p\sigma }(x)$. The subindexes $\mathit{a}$ and $\mathit{b}$
in (\ref{57}) represent the number $n$, given in Eqs. (\ref{36a}) and (\ref
{9}) respectively. In (\ref{56}) the subindexes $\mathit{a}$ and $\mathit{b}$
were already written in terms of the Landau levels for the W-bosons ($%
\mathit{m}$) and electrons ($\mathit{l}$) with the help of Eqs. (\ref{36a})
and (\ref{9}).

Note that, differently from the approach used in previous works \cite{Olivo}$%
^{,}$\cite{Grasso}$^{,}$\cite{Valle}$,$ the interaction between the magnetic
field and the W-bosons is kept in (\ref{56}) in the poles of the self-energy
operator through the effective momentum $\overline{k}^{2}$.

Expression (\ref{56}) is the general formula for the one-loop neutrino
self-energy in a constant magnetic field of arbitrary strength in the Ritus'
approach.

\subsection{Strong field approximation}

From now on, we assume that the magnetic field strength is in the range $%
m_{e}^{2}\ll eB\ll M_{W}^{2}$. Since in this case the gap between the
electron Landau levels is larger than the electron mass square ( $eB\gg
m_{e}^{2}$), it is consistent to use the LLL approximation for the electron (%
$\mathit{l}=0$). On the other hand, it is obvious that such an approximation
is not valid for the W-bosons, so for them we are bound to maintain the sum
in all Landau levels. In this approximation we have

\[
\left( 2\pi \right) ^{4}\delta ^{(4)}(p-p^{\prime })\Sigma (p)=\frac{-ig^{2}%
}{2}\int d^{4}xd^{4}y\int d^{4}q%
\sum_m\hspace{-0.5cm}\int%
d^{4}k\frac{e^{-i(p.x-p^{\prime }.y)}}{(\overline{q}^{2}+m_{e}^{2})(%
\overline{k}^{2}+M_{W}^{2})} 
\]

\[
\left\{ \overline{q}_{\shortparallel }\cdot \gamma ^{\shortparallel }\left(
I_{\mathit{m}-2,\mathit{0}}(x)I_{\mathit{m}-2,\mathit{0}}^{*}(y)+I_{\mathit{m%
},\mathit{0}}(x)I_{\mathit{m},\mathit{0}}^{*}(y)\right) \Delta \left(
-\right) L\right. 
\]

\begin{equation}
+\left. \overline{q}_{\mu }\epsilon ^{1\mu 2\nu }\gamma _{\nu }\gamma
^{5}\left( I_{\mathit{m}-2,\mathit{0}}(x)I_{\mathit{m}-2,\mathit{0}%
}^{*}(y)-I_{\mathit{m},\mathit{0}}(x)I_{\mathit{m,0}}^{*}(y)\right) \Delta
\left( -\right) L\right\}  \label{57a}
\end{equation}

To perform the integrals in $x$ and $y$ in (\ref{57a}) we should take into
account the formulas \cite{Ritus-Book}$^{,}$\cite{Ng}

\begin{equation}
\int d^{4}xe^{-ip.x}I_{m^{\prime }l^{\prime }}(x)=\frac{(2\pi )^{4}}{\sqrt{%
l^{\prime }!}\sqrt{m^{\prime }!}}\delta ^{3}(k+q-p)e^{-\widehat{p}_{\perp
}^{2}/2}e^{-ip_{1}\frac{q_{2}-k_{2}}{2eB}}e^{-isgn(eB)\left[ (l^{\prime
}-m^{\prime })\varphi \right] }J_{m^{\prime }l^{\prime }}^{*}(\widehat{p}%
_{\perp })  \label{58}
\end{equation}

\begin{equation}
\int d^{4}ye^{ip.y}I_{m^{\prime }l^{\prime }}^{*}(y)=\frac{(2\pi )^{4}}{%
\sqrt{l!}\sqrt{m!}}\delta ^{3}(k+q-p)e^{-\widehat{p}_{\perp }^{2}/2}e^{ip_{1}%
\frac{q_{2}-k_{2}}{2eB}}e^{isgn(eB)\left[ (l^{\prime }-m^{\prime })\varphi
\right] }J_{m^{\prime }l^{\prime }}(\widehat{p}_{\perp })  \label{59}
\end{equation}
where

\begin{equation}
\widehat{p}_{\perp }\equiv \sqrt{\widehat{p}_{1}+\widehat{p}_{2}},\qquad
\varphi \equiv \arctan (\widehat{p}_{2}/\widehat{p}_{1}),\qquad \widehat{p}%
_{\mu }\equiv \frac{p_{\mu }\sqrt{2\left| eB\right| }}{2eB}  \label{60}
\end{equation}

\begin{equation}
J_{m^{\prime }l^{\prime }}(\widehat{p}_{\perp })=\sum\limits_{j=0}^{\min
(l^{\prime },m)}\frac{m^{\prime }!l^{\prime }!}{j!(l^{\prime }-j)!(m^{\prime
}-j)!}\left[ isgn(eB)\widehat{p}_{\perp }\right] ^{m^{\prime }+l^{\prime
}-2j}  \label{61}
\end{equation}

Thanks to the factor $e^{-\widehat{p}_{\perp }^{2}/2}$, the contributions
from large values of $\widehat{p}_{\perp }$ are exponentially suppressed in
the electron LLL approximation. Thus, it is consistent to keep only the
smallest power of $\widehat{p}_{\perp }$ in $J_{m^{\prime }0}(\widehat{p}%
_{\perp })$ so that

\begin{equation}
J_{m^{\prime }0}(\widehat{p}_{\perp })\simeq \delta _{m^{\prime },0}
\label{62}
\end{equation}

After using Eqs. (\ref{58}), (\ref{59}) and (\ref{62}) in (\ref{57a}), and
integrating in $k$, one obtains for the neutrino self-energy in the electron
LLL approximation

\[
\Sigma (p)=-ig^{2}\pi \left| eB\right| \sum\limits_{m}\int \frac{%
d^{2}q_{\shortparallel }}{(4\pi )^{2}}\frac{1}{(q_{\shortparallel
}^{2}+m_{e}^{2})(\overline{q_{_{m}}-p}^{2}+M_{W}^{2})} 
\]

\begin{equation}
\left\{ \overline{q}_{\shortparallel }.\gamma ^{\shortparallel }(\delta
_{m,2}\delta _{m,2}+\delta _{m,0}\delta _{m,0})\Delta \left( -\right) L+%
\overline{q}_{\mu }\epsilon ^{1\mu 2\nu }\gamma _{\nu }\gamma ^{5}(\delta
_{m,2}\delta _{m,2}-\delta _{m,0}\delta _{m,0})\Delta \left( -\right)
L\right\}  \label{63}
\end{equation}
where

\begin{equation}
\overline{q_{_{m}}-p}^{2}=(q_{\shortparallel }-p_{\shortparallel
})^{2}+2(m-1/2)eB  \label{64}
\end{equation}

Note that in this approximation the sum in the W-boson Landau levels is
effectively reduced to the contribution of the two levels $m=0,2$.

Let us perform now the integration in the parallel momenta $%
q_{\shortparallel }$. With this aim, we can use the Feynman parametrization
to represent, after Wick rotation to Euclidean space and some variable
changes, the integral in (\ref{63}) as

\begin{equation}
p_{\shortparallel }\int\limits_{0}^{1}zdz\int d_{E}^{2}q_{\shortparallel }%
\frac{1}{\left( q_{\shortparallel }^{2}+\mathcal{M}^{2}\right)^{2} }
\label{65}
\end{equation}
where

\begin{equation}
\mathcal{M}^{2}=m_{e}^{2}+\left[ M_{W}^{2}+2(m-1/2)eB-m_{e}^{2}\right]
z+p_{\shortparallel }^{2}\left( 1-z\right) z  \label{66}
\end{equation}

Then, performing the integrations in $z$ and $q_{\shortparallel }$ and
taking explicitly the sum in $m$, we arrive at

\begin{equation}
\Sigma (p)=\frac{g^{2}}{(4\pi )^{2}}\left[ (\lambda -\lambda _{1}^{\prime
})p_{\shortparallel }.\gamma ^{\shortparallel }+\lambda _{2}^{\prime }p_{\mu
}\epsilon ^{1\mu 2\nu }\gamma _{\nu }\gamma ^{5}\right] \Delta \left(
-\right) L  \label{66a}
\end{equation}
where

\begin{equation}
\lambda =\frac{\left| eB\right| }{M_{W}^{2}},\qquad \lambda _{1}^{\prime }=%
\frac{m_{e}^{2}\left| eB\right| }{M_{W}^{4}}\ln (\frac{M_{W}^{2}}{m_{e}^{2}}%
)+\frac{\left| eB\right| -m_{e}^{2}}{M_{W}^{4}}\left| eB\right| ,\qquad
\lambda _{2}^{\prime }=-\frac{4\left| eB\right| ^{2}}{M_{W}^{4}}  \label{66b}
\end{equation}

The expression (\ref{66a}) can be rewritten in a more convenient way using
the following covariant form

\begin{equation}
\Sigma (p)=\left[ a_{1}p\llap/_{\shortparallel }+a_{2}p\llap/_{\bot }+bu%
\llap / +c\widehat{B}\llap / \right] L  \label{67}
\end{equation}

In (\ref{67}) $u_{\mu }$ is the four-velocity of the center of mass of the
magnetized system (background) and $\widehat{B}_{\mu }=B_{\mu }/\left|
B_{\mu }\right| $, where $B_{\mu }$ is the magnetic field in covariant
notation $B_{\mu }=\frac{1}{2}\epsilon _{\mu \nu \rho \lambda }u^{\nu
}F^{\rho \lambda }$. Notice that when a magnetic field is present, to form
the structure of $\Sigma $ we have to consider, in addition to the usual
tensors $p_{\mu }$, $g_{\mu \nu }$, and $\epsilon _{\mu \nu \rho \lambda }$,
the vectors $B_{\mu }$ and $u_{\mu }$. The four-velocity vector $u_{\mu }$
can be introduced in this case because the presence of a constant magnetic
field picks up a special Lorentz frame: the rest frame (on which $u_{\mu
}=(1,0,0,0)$) where the magnetic field is defined ($u_{\mu }F^{\mu \nu }=0$).

In a trivial vacuum, $\Sigma $ would depend only on the four-momentum $%
p_{\mu }$ showing no difference between longitudinal and transverse
components $(a_{1}=a_{2})$ in agreement with the Lorentz invariance of the
system. However, when a nontrivial background is present, the structure of $%
\Sigma $ is enriched with new terms related to the symmetries broken by the
background. In the present situation, since the magnetic field introduces a
special Lorentz frame, the four-velocity $u_{\mu }$ is needed to rewrite the
structure in a covariant way, a situation similar to the finite temperature
case\cite{Fradkin}$.$ Moreover, because of its special direction in the
3-dimensional space, the magnetic field breaks one more symmetry: the O(3)
rotational symmetry. This new symmetry breaking is responsible for the
appearance of the structure associated to the unit vector $\widehat{B}$.

Notice that the separation between longitudinal and transverse momenta
contributions in $\Sigma $ (i.e. $a_{1}\neq a_{2}$ in (\ref{67})), an effect
normally occurring in the presence of a constant magnetic field, has also a
covariant representation in terms of the basic tensors of the problem,

\begin{equation}
p\llap/_{\shortparallel }=p^{\nu }W_{\nu \rho }W^{\mu \rho }\gamma _{\mu }
\label{67a}
\end{equation}

\begin{equation}
p\llap/_{\bot }=p^{\nu }\widetilde{W}_{\nu \rho }\widetilde{W}^{\mu \rho
}\gamma _{\mu }  \label{67b}
\end{equation}
where

\begin{equation}
W_{\nu \rho }=(u^{\alpha }\widehat{B}^{\beta }-u^{\beta }\widehat{B}^{\alpha
})  \label{67c}
\end{equation}

\begin{equation}
\widetilde{W}_{\nu \rho }=\frac{1}{2}\epsilon _{\nu \rho \alpha \beta
}W^{\alpha \beta }  \label{67d}
\end{equation}

The coefficients $a_{1}$, $a_{2}$, $b$, and $c$ in (\ref{67}) are Lorentz
scalars that depend on the parameters of the theory and the used
approximation. From (\ref{66a})-(\ref{66b}) one can see that in the
strong-field approximation here considered their leading contributions in
powers of $1/M_{W}^{2}$ are given by

\begin{equation}
a_{1}=\frac{g^{2}}{(4\pi )^{2}}\lambda ,\qquad a_{2}=0,\qquad b=a_{1}\chi
_{p},\qquad c=a_{1}\omega _{p}  \label{68}
\end{equation}
with

\begin{equation}
\omega _{p}=p\cdot u,\qquad \chi _{p}=p\cdot \widehat{B}  \label{69}
\end{equation}

It is clear from the above equations that the longitudinal and transverse
neutrino modes of propagation behave quite differently. This means that the
strong magnetic field gives rise to an anisotropy in the neutrino
propagation that is reflected in a neutrino self-energy mainly depending on
the spatial momentum parallel to the applied field (\ref{67}), (\ref{68}).
We point out that in the weak-field approximation a neutrino anisotropic
propagation was found in Ref [11], while in Ref. [12] the splitting,
characteristic in the presence of a magnetic field, between longitudinal and
transverse momentum components, was absent. The last result was a
consequence of the mass shell condition for vacuum $\gamma \cdot p=0$ that
was imposed through out the calculation in [12]. Moreover, it should be
notice that no linear term in $B$ was found in [12], contrary to the
behavior reported in [11], and to the one we found in $\lambda $ and $%
\lambda ^{\prime }$ from Eqs. (\ref{68})-(\ref{69}).

For the case of neutrino propagation in a magnetized medium ($\mu \neq 0$),
a self-energy structure similar to (\ref{67}) has been reported \cite{Olivo}$%
.$ However, in that case$,$ the coefficients $b$ and $c$ are proportional to
the difference between the electron and positron densities which are
functions of the electron chemical potential.

\section{Neutrino dispersion relation and index of refraction at strong
magnetic field}

Using the results (\ref{67}), (\ref{68}) in the dispersion equation for
neutrinos propagating in the external magnetic field

\begin{equation}
\det \left[ \gamma \cdot p-\Sigma (p)\right] =0  \label{70}
\end{equation}
one obtains the following solution

\begin{equation}
\omega _{p}\simeq \left| \overrightarrow{p}\right| (\pm 1+a_{1}\sin
^{2}\alpha ).  \label{71}
\end{equation}
In Eq. (\ref{71}), $\alpha $ is the angle between the direction of the
neutrino momentum and that of the applied magnetic field. Positive and
negative signs correspond to neutrino and antineutrino energies respectively.

To obtain the neutrino index of refraction $n$, we substitute (\ref{71})
into the formula

\begin{equation}
n\equiv \frac{\left| \overrightarrow{p}\right| }{\omega _{p}}  \label{72}
\end{equation}
to find

\begin{equation}
n\simeq 1-a_{1}\sin ^{2}\alpha  \label{73}
\end{equation}

From Eqs. (\ref{71}) and (\ref{73}) it is clear that neutrinos moving with
different directions in the magnetized space will have different dispersion
relation and consequently, different index of refraction. That is, although
the neutrinos are electrically neutral, the magnetic field, through quantum
corrections, can produce anisotropic neutrino propagation. The order of the
asymmetry is $g^{2}\frac{\left| eB\right| }{M_{W}^{2}}$.

An asymmetric neutrino propagation depending on the difference between the
number densities of electrons and positrons was previously found in a
charged medium ($\mu \neq 0$) \cite{Olivo}$.$ There, the asymmetric term
changes its sign when the neutrino reverses its motion. This property was
suggested\cite{Kusenko} to be the cause of the peculiarly high velocities
observed in pulsars$.$ In our case, however, the asymmetric term in the
dispersion relation (\ref{71}) does not change its sign by changing $\alpha $
by $-\alpha $. On the other hand, it is clear from (\ref{73}) that neutrinos
moving along the external magnetic field $(\alpha =0)$ have index of
refraction similar to the one for the free case $(n=1)$, while the index of
refraction for neutrinos moving perpendicularly to the direction of the
magnetic field $(\alpha =\pi /2),$ has a maximum departure from the
free-case value.

\section{Final remarks and cosmological applications}

In this paper we have found the vacuum contribution ($T=0,$ $\mu =0$) of the
neutrino self-energy in the strong-field regime ($m_{e}^{2}\ll eB\ll
M_{W}^{2}$). The obtained self-energy depends only on the longitudinal
neutrino momentum $p_{\Vert }$. This fact is responsible of the strongly
anisotropic neutrino propagation discussed in Sec. IV. Nevertheless,
contrary to what occurs in the case with $\mu \neq 0$\cite{Olivo}$,$ the
asymmetric term in the dispersion relation maintains the sign when the
neutrino reverses its motion. From (\ref{66b}) we can see that in our
approximation the terms in $\Sigma $ depending on the charged lepton mass $%
m_{e}$ are negligible small ($1/M_{W}^{4}$).

Our results may find applications in the physics of neutrinos in the early
Universe. First of all, notice that the existence of strong magnetic fields
in the early Universe seems to be a very plausible idea\cite{enq}$,$ since
they may be required to explain the observed galactic magnetic fields, $%
B\sim 2\times 10^{-6}$ $G$ on scales of the order of $100$ $kpc$ \cite
{Galaxies}$.$

The strength of the primordial magnetic field in the neutrino decoupling era
can be estimated from the following reasoning. Based on constraints derived
from the successful nucleosynthesis prediction of primordial $^{4}He$
abundance \cite{He}$,$ as well as on the neutrino mass and oscillation
limits, an upper bound for the magnetic field produced in the early Universe
prior to primordial nucleosynthesis \cite{Enqvist} has been predicted. A
formula for the upper bound at the QCD phase transition is \cite{Enqvist}

\begin{equation}
B_{QCD}\lesssim \frac{10^{21}G}{\sum\limits_{i}m_{\nu _{i}}/eV}  \label{74}
\end{equation}
Taking into account the cosmological constraint on the sum of stable
neutrino masses $\sum\limits_{i}m_{\nu _{i}}\lesssim 100$ $eV$, the relation
(\ref{74}) implies that at $T_{QCD}=200$ $MeV$ the estimated upper limit for
the primordial magnetic field is 
\begin{equation}
B_{QCD}\lesssim 10^{19}G  \label{75}
\end{equation}

On the other hand, taking into account the magnetic field effect of
increasing $n\leftrightarrow p$ reaction rates in primordial
nucleosynthesis, the magnetic field upper limit at the end of
nucleosynthesis ($T=10^{9}$ $K$) is \cite{NS}

\begin{equation}
B_{NS}\lesssim 10^{11}G  \label{75a}
\end{equation}

These values are in agreement with the equipartition principle that states
that the magnetic energy can only be a small fraction of the Universe energy
density. This argument leads to the relation $B/T^{2}\sim 2$.

Therefore, it is reasonable to assume that between the QCD phase transition
epoch and the end of nucleosynthesis a primordial magnetic field in the range

\begin{equation}
m_{e}^{2}\leq eB\leq M_{W}^{2}  \label{76}
\end{equation}
could have been present.

On the other hand, the early Universe, unlike the dense stellar medium, is
almost charge symmetric ($\mu =0$), since the particle-antiparticle
asymmetry in the Universe is believed to be at the level of $%
10^{-10}-10^{-9} $, while in stellar material is of order one. Thus, to
investigate neutrino propagation in cosmology, we would have to consider, in
addition to the vacuum contribution, $\Sigma _{\nu }^{T=0}\left(
m_{e},M_{W},eB\right) $, of the neutrino self-energy, the $\mu =0$ thermal
contribution, $\Sigma _{\nu }^{T\neq 0}\left( m_{e},M_{W},eB,T\right) $,

\begin{equation}
\Sigma _{\nu }\left( m_{e},M_{W},eB,T\right) =\Sigma _{\nu }^{T=0}\left(
m_{e},M_{W},eB\right) +\Sigma _{\nu }^{T\neq 0}\left( m_{e},M_{W},eB,T\right)
\label{77}
\end{equation}

Notice that, when dealing with possible cosmological applications of our
results, the vacuum neutrino self-energy calculated in Sec. III using the
LLL approximation of the fermion Green function should be taken as a
qualitative result. To understand this, we recall that a reasonable
primordial magnetic field in the neutrino decoupling era would satisfy

\begin{equation}
m_{e}^{2}\ll eB\sim 2T^{2}\qquad  \label{82}
\end{equation}

For such fields, the effective gap between the Landau levels is $%
eB/T^{2}\sim \mathcal{O}(1)$. Clearly, in this case the weak-field
approximation, which would correspond to a high-temperature approximation,
cannot be used because field and temperature are comparable. On the other
hand, because the thermal energy is of the same order of the energy gap
between Landau levels, it is barely enough to induce the occupation of only
a few of the lower Landau levels. Therefore, the LLL approximation, even
though too radical here since, strictly speaking, this is not a clear-cut
strong-field case, it will provide a qualitative description of the neutrino
propagation. In other words, the vacuum structure associated to the
anisotropic propagation should still be present for fields $eB\sim 2T^{2}$,
although the coefficient $c$ in Eq. (\ref{67}) may be quantitatively
different.

A more quantitative treatment of the neutrino propagation in a field $eB\sim
2T^{2}$ would require numerical calculations due to the lack of a leading
parameter. In this sense, the extension to spin-one charged particles of the
Ritus' method developed in Section II can be very useful, since expression (%
\ref{56}) is a suitable representation to be used to numerically find the
coefficients $a$, $b$ and $c$ of the general structure of the self-energy (%
\ref{67}).

Therefore, one can expect that the anisotropic propagation of neutrinos at
zero chemical potential, found in our calculations at the LLL approximation,
would be reflected in the propagation of neutrinos in the early universe if
primordial fields satisfying the condition (\ref{82}) were present. Even if
the anisotropic term results small, it would account for a qualitatively new
effect. If that is the case, one can envision that the anisotropy would
leave a footprint in a yet to be observed relic neutrino cosmic background.
If such an effect were detected, it would provide a direct experimental
proof of the existence of strong magnetic fields in the early Universe.

\textbf{Acknowledgment}

We thank V. Gusynin and J. F. Nieves for enlightening discussions. Ferrer
and Incera would like to thank the Institute for Space Studies of Catalonia
and the Department of Structure and Constituents of Matter of the University
of Barcelona for their warm hospitality during the time this work was done.
This work has been supported in part by NSF grants PHY-0070986 and
PHY-9722059(EF and VI), NSF POWRE grant PHY-9973708 (VI) and by grants
PB96-0925 DGICYT and 1999SGR-00257 CIRIT (EE).


\begin{thebibliography}{99}
\bibitem{footnote1}  Although the calculations are specialized to the
Feynman gauge, the neutrino energy-momentum relation, found below, will be
of course gauge independent, since it is obtained from the pole of the
neutrino propagator.
\end{thebibliography}
\end{document}